\documentclass[letter]{aa}
\usepackage[utf8]{inputenc}
\usepackage{threeparttable,pifont, url}
\usepackage{natbib}
\usepackage{graphicx,caption,subcaption}

\usepackage{graphicx}   
\usepackage{amsmath}    
\usepackage{amssymb}    

\newcommand{\bi}{\begin{itemize}}
\newcommand{\ei}{\end{itemize}}
\newcommand{\bcs}{\begin{columns}}
\newcommand{\ecs}{\end{columns}}
\newcommand{\ec}{\end{column}}
\newcommand{\bc}{\begin{column}}

\newcommand{\tiu}{\ensuremath{\mathrm{J}~\mathrm{m}^{-2}~\mathrm{s}^{-0.5}~\mathrm{K}^{-1}}}

\begin{document}

\title{(3200)~Phaethon: Bulk density from Yarkovsky drift detection}

\titlerunning{(3200) Phaethon: Bulk density from Yarkovsky drift detection}

\author{
        J.~Hanu\v{s}\inst{1}
        \and
        D.~Vokrouhlick\'y\inst{1}         
        \and         
        M.~Delbo'\inst{2}
        \and
        D.~Farnocchia\inst{3}
        \and
        D.~Polishook\inst{4}
        \and
        P.~Pravec\inst{5}
        \and
        K.~Hornoch\inst{5}
        \and
        H.~Ku\v{c}\'akov\'a\inst{5}
        \and
        P.~Ku\v{s}nir\'ak\inst{5}
        \and
        R.~Stephens\inst{6}
        \and
        B.~Warner\inst{7}
}

   \institute{
        Institute of Astronomy, Charles University, Prague, V Hole\v sovi\v ck\'ach 2, CZ-18000, Prague 8, Czech Republic
        \and
        Université Côte d'Azur, Observatoire de la Côte d'Azur, CNRS, Laboratoire Lagrange, France
        \and
        Jet Propulsion Laboratory, California Institute of Technology, Pasadena, CA 91109, USA 
        \and
        Department of Earth and Planetary Sciences, Weizmann Institute of Science, Rehovot 0076100, Israel 
        \and
        Astronomical Institute, Academy of Sciences of the Czech Republic, Fri\v cova 1, CZ-25165 Ond\v rejov, Czech Republic
        \and
        Center for Solar System Studies, 11355 Mount Johnson Court, Rancho Cucamonga, CA 91737 USA
        \and
        Center for Solar System Studies, 446 Sycamore Ave., Eaton, CO 80615, USA
}

\date{July 2018}

\abstract
{The recent close approach of the near-Earth asteroid (3200)~Phaethon offered a rare
 opportunity to obtain high-quality observational data of various types.}
{We used the newly obtained optical light curves to improve the spin and shape model of
 Phaethon and to determine its surface physical properties derived by thermophysical
 modeling. We also used the available astrometric observations of Phaethon, including those
 obtained by the Arecibo radar and the Gaia spacecraft, to constrain the secular drift
 of the orbital semimajor axis. This constraint allowed us to estimate the bulk density by assuming
 that the drift is dominated by the Yarkovsky effect.}
{We used the convex inversion model to derive the spin orientation and 3D shape model
 of Phaethon, and a detailed numerical approach for an accurate analysis of the Yarkovsky
 effect.}
{We obtained a unique solution for Phaethon's pole orientation at $(318\degr,-47\degr)$ 
 ecliptic longitude and latitude (both with an uncertainty of $5\degr$), and
 confirm the previously reported thermophysical properties ($D=5.1\pm0.2$~km,
 $\Gamma=600\pm 200$ \tiu). Phaethon has a top-like shape with possible north-south
 asymmetry. The characteristic size of the regolith grains is $1-2$~cm. The orbit analysis reveals a
 secular drift of the semimajor axis of $-(6.9\pm 1.9)\times 10^{-4}$ au~Myr$^{-1}$. 
 With the derived volume-equivalent size of 5.1~km, the bulk density is
 $1.67\pm 0.47$ g~cm$^{-3}$. If the size is slightly larger  $\sim 5.7-5.8$~km, as suggested
 by radar data, the bulk density would decrease to $1.48\pm 0.42$ g~cm$^{-3}$. We further investigated the suggestion that Phaethon may be in a cluster with asteroids (155140)~2005~UD and (225416)~1999~YC that was formed by rotational fission of a critically spinning parent body.}
{Phaethon's bulk density is consistent with typical values for large ($>100$~km) C-complex
 asteroids and supports its association with asteroid (2)~Pallas, as first suggested by
 dynamical modeling. These findings render a cometary origin unlikely for Phaethon.}

\keywords{minor planets, asteroids, individual: (3200) Phaethon, methods:numerical, methods: observational, astrometry, celestial mechanics}

\maketitle

\section{Introduction}

Physical properties of the low-perihelion near-Earth asteroid Phaethon, target of the
proposed JAXA DESTINY$^+$ mission \citep{Arai2018}, have so far been inferred from a wide range of
datasets: photometry in optical \citep{Krugly2002,Ansdell2014} and infrared bands
\citep{Green1985,Tedesco2004}, spectroscopy \citep{Licandro2007, Hanus2016b}, polarimetry
\citep{Devogele2018,Ito2018}, and radar \citep{Taylor2018}. Other studies were dedicated to Phaethon's comet-like activity
\citep{Jewitt2010, Jewitt2013, Li2013, Ye2018}, to the associated Geminid meteor stream
\citep{Gustafson1989, Williams1993, TrigoRodriguez2004}, or to the dynamical link with
the main-belt asteroid (2)~Pallas \citep{deLeon2010, Todorovic2018}. 

Despite these numerous studies, the true nature of
Phaethon has not yet been convincingly revealed. There is even a controversy involving the
basic physical properties of Phaethon, such as its size and geometric visible albedo:
recent studies based on polarimetric observations \citep{Devogele2018,Ito2018} report
a significantly lower geometric albedo than has been inferred from thermal infrared data.
Similarly, a possibly larger size than previous diameter determinations from thermal infrared observations was reported from the analysis of delay-Doppler observations
by \citet{Taylor2018}. These persisting inconsistencies motivated us to apply independent
methods and improve our previous thermophysical modeling to ultimately understand the
nature and origin of this intriguing object.

The polarimetric and dynamical studies mentioned above are in favor of a physical link between Phaethon
and Pallas. In their view, Phaethon is an escapee member from the Pallas collisional
family. This association is also supported by the spectroscopic studies in the visible and
near-infrared \citep{Licandro2007, deLeon2012}. On the other hand, some authors remain
in favor of a cometary origin for Phaethon \citep[e.g.,][]{TrigoRodriguez2004,Borovicka2010}.
In order to shed light on this fundamental issue, we first determine a unique model
of Phaethon's spin state and its shape from a thermophysical model. Next, we use all
available astrometric data to prove that accurate orbit determination requires that the Yarkovsky effect is included, which results in a steady decrease of the semimajor axis
\citep[e.g.,][]{Vokrouhlicky2015}. Because of its non-gravitational origin, a detailed
theoretical model of the Yarkovsky effect, if fed by our spin and shape solution,
allows us to constrain Phaethon's bulk density. This parameter helps us to infer its internal composition and conclude about its origin.

\section{Astrometric observations and orbit determination}\label{sec:astrometry}

Astrometric data exist since the discovery of Phaethon by the Infrared Astronomical Satellite in October 1983 \citep{Green1983}.
As of July~2018, 4782 astrometric observations have been reported to the Minor Planet Center by ground-based
observatories and 28 observations by the WISE spacecraft.%
\footnote{\url{https://www.minorplanetcenter.net/db_search/show_object?utf8=\%E2\%9C\%93&object_id=3200}}
To this dataset of optical astrometry, we applied the \citet{Farnocchia2015a} star catalog
debiasing and the \citet{Veres2017} weighting scheme. Isolated observations that showed
localized biases or internal inconsistencies were de-weighted or excluded from the fit.

In addition to the ground-based observations and WISE data, the Gaia spacecraft observed
Phaethon during 12 transits between September 2014 and February 2016. The
corresponding astrometry was part of the Gaia DR2 release \citep{Spoto2018}. The Gaia
small-body astrometry is decoupled into two components, along scan (AL), and across scan (AC).
These two components are only weakly correlated; the AL component has a much greater accuracy
than the AC component. Nevertheless, we made use of the full correlated
observation error covariance model and found that the characteristic uncertainties were
$\simeq 10$~mas in the AL direction and $\simeq 0.6$\arcsec ~in the AC direction. To avoid
problems with transit-specific systematic errors, we selected a single observation for
each of the 12 transits.

Because of its low Earth MOID of $\simeq 0.02$~au, Phaethon occasionally experiences
close approaches to the Earth. During the 2007 and 2017 approaches, a total of six 
delay measurements were collected from the Arecibo and Goldstone radars \citep{Taylor2018}.%
\footnote{\url{https://ssd.jpl.nasa.gov/?grp=ast&fmt=html&radar=}}
This wealth of observational data places extremely tight constraints on the orbit of Phaethon.
For instance, the formal uncertainty in semimajor axis is only $\simeq 64$~m and only 20 ms
in orbital period.

We used the aforementioned astrometric observations for orbital determination: Our force model includes the Newtonian gravity of the Sun, the planets, Pluto, Moon, the 16 largest perturbers in the main belt, and relativistic effects \citep[e.g.,][]{Farnocchia2015b}.
Since Phaethon has a low perihelion and experiences close encounters to the Earth, we
also included perturbations that are due the oblateness of the Sun and the Earth. To fit the complete
dataset, and in particular both the 2007 and 2017 radar apparitions, it was also necessary
to include non-gravitational perturbations, in particular, the Yarkovsky effect. We describe
Yarkovsky modeling and its parametric dependence in Sect.~\ref{sec:yarko}.
\begin{figure}
 \begin{center}
  \resizebox{\hsize}{!}{\includegraphics{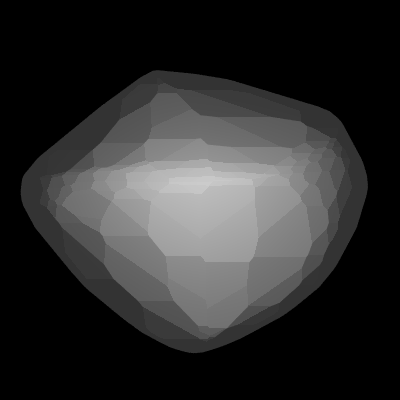}\includegraphics{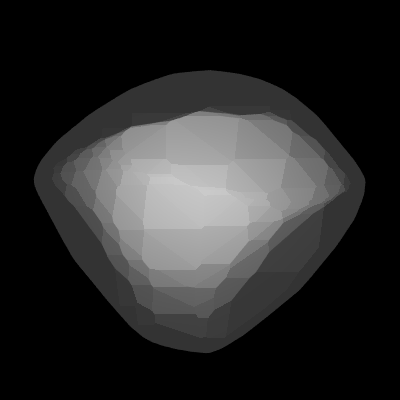}\includegraphics{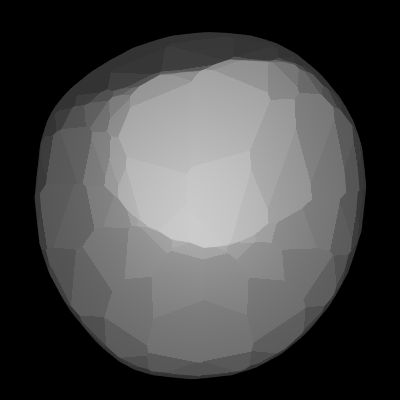}}\\
 \end{center}
 \caption{\label{fig:shape} Revised shape model of Phaethon. Three different viewing
  geometries are shown: the left and middle images are equator-on views rotated by 90$^{\circ}$,
  the right image is a pole-on view.}
\end{figure}

\section{Results}

\subsection{Revised spin and shape model}\label{sec:shape}

We applied the convex inversion method of \citet{Kaasalainen2001b} and
\citet{Kaasalainen2001a} to the optical dataset of 70 light curves (described in
Appendix~\ref{sec:lcs} and listed in Table~\ref{tab:lcs})
following exactly the procedure of \citet{Hanus2016b}. Specifically, we scanned
rotation periods in the proximity of the expected value while testing ten initial pole
solutions for each sampled period. Four poles were selected on the equator with 90 degrees difference in longitude, and three poles in each hemisphere with the latitude $\pm60$ degrees and with 120 degrees difference in longitude. We assumed that all solutions within a $3\sigma$ uncertainty interval had $\chi^2<(1+3\sqrt{2/\nu})\,\chi^{2}_\textrm{min}$, where $\chi^{2}_\textrm{min}$ is the $\chi^2$ of the best-fitting solution and $\nu$ is the number of degrees of freedom. This threshold to consider the solution acceptable was used before in \citet{Vokrouhlicky2017} or \citet{Durech2018a}\footnote{Be aware of the typo in the referenced equation in these two studies.} and corresponded to a $\sim7$\% increase in $\chi^{2}_\textrm{min}$ value. Only the best-fitting solution fulfilled the $3\sigma$ condition on the $\chi^2$. To further verify that the best-fitting solution was the only one acceptable, we also visually inspected the light-curve fit with the second-best-fitting period, similarly as in \citet{Hanus2016b}, see their Figs.~3 and 4 for illustration. This solution was already inconsistent with several individual light-curves. Therefore, we considered the difference in $\chi^2$ as significant and rejected all periods except for the best-fitting one. Next, we ran the convex inversion with the unique period and multiple pole orientations (isotropically distributed on a sphere with a 30 degree difference) as starting points of the optimization procedure and derived a single solution within the $3\sigma$ uncertainty interval defined above. Again, we visually inspected the light-curve fit with the second-best-fitting pole orientation and rejected this solution and also considered all other solutions as non-acceptable. The final solution is given in Table~\ref{tab:inversion} together with the previous determinations. It is notable that our analysis and the recent study of \citet{Kim2018} provide for the first time a
unique shape and spin solution that is consistent with the preferred solution of \citet{Hanus2016b}.
There are two differences between our old and revised models: (i) the relative dimension along the
rotation axis (or the $c/a$ ratio) is now smaller by $\sim$10\%, which is expected
because this dimension is generally the least constrained by the optical data, and
(ii)~the pole directions are about $8$ degrees apart.

We find that the overall shape of Phaethon is nearly axially symmetric: the x-y projection
is not far from a circle with $b/a$ of $\sim 0.94$ (right panel of Fig.~\ref{fig:shape}).
Moreover, there seems to be a hint of an equatorial ridge and the top-shape like appearance \citep[also noted by][from radar observations]{Taylor2018}
that is often found in the sub-kilometer and kilometer-sized fast rotators
\citep{Ostro2006, Busch2011, Naidu2015}. Additionally, our model suggests a north-south
asymmetry of Phaethon, with the northern hemisphere slightly suppressed. Interestingly, our
model for Phaethon is reminiscent of that of the Hayabusa~2 mission target (162173)~Ryugu,
whose recent public images revealed its top-shape appearance \citep{Hasegawa2008}.
\begin{table*}
 \caption{\label{tab:inversion}Rotation state parameters derived for Phaethon from different
  photometric datasets. The table gives the ecliptic longitude $\lambda$ and latitude $\beta$
  of all possible pole solutions, the sidereal rotation period $P$, and the reference.}
\centering
\begin{tabular}{cccc c c} \hline
\multicolumn{1}{c} {$\lambda_1$} & \multicolumn{1}{c} {$\beta_1$} & \multicolumn{1}{c} {$\lambda_2$} & \multicolumn{1}{c} {$\beta_2$} & \multicolumn{1}{c} {$P$} & Note \\
\multicolumn{1}{c} {[deg]} & \multicolumn{1}{c} {[deg]} & \multicolumn{1}{c} {[deg]} & \multicolumn{1}{c} {[deg]} & \multicolumn{1}{c} {[hours]} &  \\
\hline\hline
 $276\pm15$ & $-15\pm15$ & $97\pm15$ & $-11\pm15$ & $3.5906\phantom{58}\pm0.0001\phantom{02}$  & \citet{Krugly2002} \\
            &            & $85\pm13$ & $-20\pm10$ & $3.6032\phantom{58}\pm0.0008\phantom{02}$  & \citet{Ansdell2014} \\
 $319\pm\phantom{1}5$    & $-39\pm\phantom{1}5$   & $84\pm\phantom{1}5$ & $-39\pm\phantom{1}5$ & $3.603958\pm0.000002$ & \citet{Hanus2016b} \\
 $308\pm10$ & $-52\pm10$  &  &  & $3.603957\pm0.000001$ & \citet{Kim2018} \\
 $318\pm\phantom{1}5$    & $-47\pm\phantom{1}5$   &  &  & $3.603957\pm0.000001$ & This work \\
\hline
\end{tabular}
\end{table*}

\subsection{Updated thermophysical properties}\label{sec:TPM}

Given the new and unique spin and shape model, we repeated the thermophysical
modeling (TPM) of \citet{Hanus2016b}. The revised TPM solution is consistent with the
previous one, mostly because the shape model derived here is similar to that of \citet{Hanus2016b}. Therefore, we
do not report in detail the new results as they are essentially identical to those
in \citet{Hanus2016b}: (i) equivalent size $D=5.1\pm 0.2$~km, (ii) geometric albedo
$p_{\mathrm{V}}=0.122\pm0.008$, and (iii) thermal inertia $\Gamma=600\pm 200$ \tiu \,(all
formal uncertainties).
\begin{figure}
 \resizebox{1.0\hsize}{!}{\includegraphics{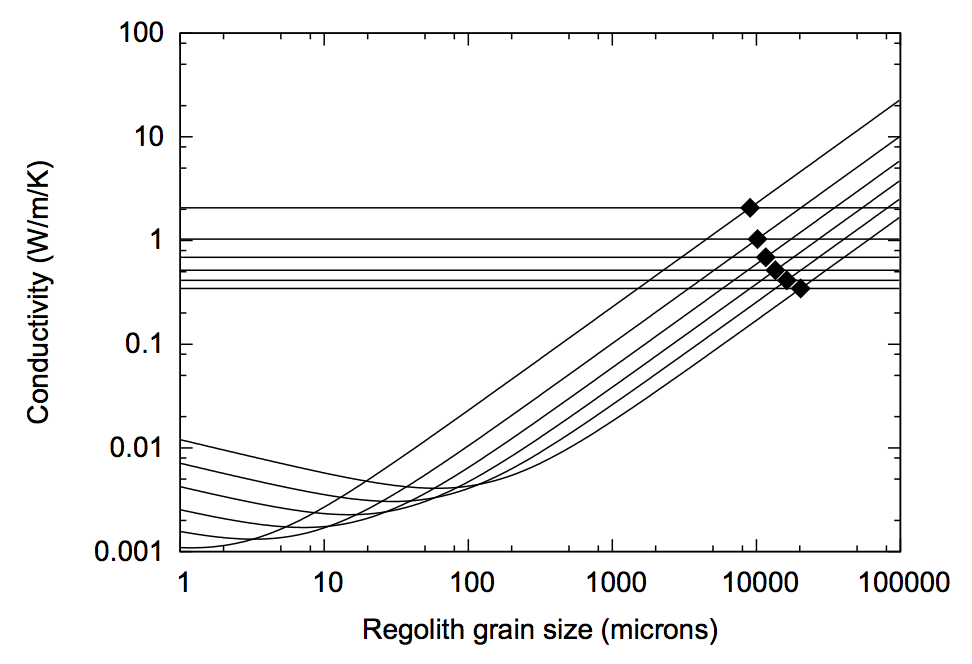}}\\
 \caption{Phaethon's regolith grain-size. Horizontal lines indicate the derived values
  of the thermal conductivity, following Eq.~(\ref{E:kappaAst}), for the different volume
  filling factors $\phi$ of the material and for the nominal thermal inertia value of
  $600$ \tiu (from top to bottom: $\phi$ = 0.6, 0.5, 0.4, 0.3, 0.2, and 0.1). The curves
  represent the thermal conductivity of a regolith with average thermophysical properties
  of CC meteorites \citep[see][]{Opeil2010} as a function of the regolith grain size again
  for the different volume filling factors $\phi$. The intersection of the curves with
  the horizontal lines (for the same $\phi$) give the inferred characteristic grain size
  of the regolith at the abscissa.}
\label{fig:grainSizePhaethon}
\end{figure}
\begin{table}
 \caption{\label{tab:grainSize}Four cases of the regolith volume filling factor $\phi$ and
  regolith grain size $r_\mathrm{g}$ combinations, and the corresponding parameters of the
  thermal conductivity dependence on the temperature $T$ given by Eq.~(\ref{eq:conductivity}).}
 \begin{center}
 \begin{tabular}{ccccc}
  \hline
   Case & $\phi$ & $r_\mathrm{g}$ & $\kappa_\mathrm{b}\,(\times 10^{-5})$ & $a\,(\times 10^{-8})$ \\
        &        &     [mm]     &  [W\,m$^{-1}$\,K$^{-1}$] &
    [W\,m$^{-1}$\,K$^{-4}$] \\
  \hline
  1 & 0.3 & 12 & $\phantom{2}8.8\pm1.4$ & $1.7\phantom{7}$ \\
  2 & 0.4 & 14 & $11\phantom{.8}\pm1\phantom{.4}$  & $1.3\phantom{7}$ \\
  3 & 0.5 & 16 & $20\phantom{.8}\pm2\phantom{.4}$  & $0.97$ \\
  4 & 0.6 & 20 & $32\phantom{.8}\pm3\phantom{.4}$  & $0.81$ \\
\hline
 \end{tabular}
 \end{center}
\end{table}

\subsection{Regolith grain size and thermal conductivity}\label{sec:grainSize}

We used the method of \citet{Gundlach2013} to determine the grain size of
the surface regolith of Phaethon. 
The method uses the asteroid thermal inertia $\Gamma=600\pm 200$ \tiu \, that was obtained by \citet{Hanus2016b} has also been
confirmed by us here to infer the thermal conductivity $\kappa$ of the regolith,
\begin{equation}
 \kappa = \frac{\Gamma^2}{\phi \rho c}, \label{E:kappaAst}
\end{equation}
where $c$ is the specific heat capacity, and $\rho$ is the grain density, as a function of the regolith grain filling factor $\phi$. These conductivity values are represented by the horizontal lines in Fig.~\ref{fig:grainSizePhaethon}.



Since the regolith filling factor is unconstrained, we
considered its values to range between 0.1 (extremely fluffy packing, which is plausible only for small regolith particles) and 0.6 (close to the densest packing of equal-sized particles) with $\Delta \phi$=0.1 step.
The values of $\rho$ and $c$ represent typical C-type thermophysical properties of
the regolith \citep[see][]{Opeil2010}. We only used the nominal $\Gamma$
value for the grain size determination. The relative uncertainty of 30\% in the value of
the thermal inertia translates into variations of the horizontal lines in
Fig.~\ref{fig:grainSizePhaethon} of about $60-70$\%, which is still within the range
encompassed by our lack of knowledge of the value of $\phi$. 

Next, the \citet{Gundlach2013} model calculates the thermal conductivity of a granular medium (the regolith) as a function of the size of the grain and temperature, assuming density, heat capacity, and thermal conductivity of the solid grains. The thermal inertia for Phaethon was derived from observations obtained at 1.1~au and 1~au from
the Sun. Using $p_\mathrm{V} = 0.122$ and $G=0.15$ from \citet{Hanus2016b}, we obtained a
value of the bolometric Bond albedo $A=0.048$. Using $\eta=1.6$ as in \citet{Harris1998},
we estimate sub-solar temperature $T_\mathrm{SS}$ of 339~K and 355~K at 1.1~au and 1~au, respectively.
We used the mean of these values, namely 347~K. Using these input parameters for the model, we now determined
the curves in Fig.~\ref{fig:grainSizePhaethon} following the model of \citet{Gundlach2013}.
The abscissa value of the intersection points between these curves and the horizontal lines of Fig.~\ref{fig:grainSizePhaethon} indicate the typical radius of the regolith grains. We find that the typical regolith grain size for Phaethon is about $1-2$~cm.

The regolith volume filling factors of $0.1$ or $0.2$ are probably unrealistic for the
larger grains derived above, so we decided to consider only the four cases listed in
Table~\ref{tab:grainSize}. Using Eq.~(5) of \citet{Gundlach2013}, we calculated
the temperature-dependent thermal conductivity $\kappa$ for these four cases for temperatures between 100 and 1300~K. For the
temperature range and regolith grain size appropriate for Phaethon, Eq.~(5) of \citet{Gundlach2013} gives $\kappa \propto T^3$ and may be approximated
with a simpler relation in the form \citep{Keihm1984,Delbo2015}
\begin{equation}\label{eq:conductivity}
 \kappa = \kappa_\mathrm{b} + a T^3, 
\end{equation}
where the first term corresponds to the solid-state thermal conductivity $\kappa_\mathrm{b}$, the second
term is due to thermal conductivity by photons, and $T$ is the temperature. By fitting Eq.~(\ref{eq:conductivity}) to the calculated values, we obtained the parameter values reported in Table~\ref{tab:grainSize}.

\subsection{Yarkovsky drift detection and bulk density}\label{sec:yarko}

Because of its extremely eccentric orbit ($e\simeq 0.9$), Phaethon presents a particularly challenging case for an accurate analysis
of the Yarkovsky effect. As a result of the temperature-dependent thermal conductivity described 
by Eq.~(\ref{eq:conductivity}), the value of this parameter changes by a factor $\simeq 20$
between perihelion (0.14 au) and $1$~au. Similarly, the thermal parameter, directly dictating
angular lag of the thermal effects \citep[e.g.,][]{Bottke2006}, may change by a
factor of $\simeq 5$ between perihelion and $1$~au, and the radiation flux changes
by a factor of $\simeq 50$ in the same range. All these large variations limit the validity of
simplified approaches to the Yarkovsky effect modeling, and warrant the adoption of a fully fledged numerical approach.

We adopted the model developed by \citet{Capek2005}, in which 1D thermal
conduction below each of the surface facets is solved numerically with the nonlinear
Robin boundary condition at the surface, and the assumption of an isothermal core at a
sufficient depth is made. A temperature-dependence of the thermal conductivity following
Eq.~(\ref{eq:conductivity}) was used. For the sake of simplicity, the specific heat capacity $c$ was
assumed constant, $c=560$ J~kg$^{-1}$~K$^{-1}$, and the regolith grain density obtained
for C-type meteorites was used, $\rho=3.11$ g~cm$^{-3}$ \citep[both from][]{Gundlach2013}.
We ran solutions for four values of the packing factor $\phi$ in the range between
$0.3$ and $0.6$. Each time, the parameters of the thermal conductivity were adjusted
to satisfy the constraints from thermal observations described in Sec.~\ref{sec:grainSize}
(see Table~\ref{tab:grainSize}). The time domain of one revolution about the Sun was
divided into steps of $60$~s, short enough when compared to the $\simeq 3.6$~hr
rotation period, and the space grid describing the depth below each of the surface increased
 exponentially, as described in \citet{Capek2005}. We ensured that at
each depth, the von Neumann stability condition was satisfied. Typically, ten iterative steps
of the algorithm provide the temperature with an accuracy of one degree or better in the
whole space and time domain of the solution. The shape and spin state of Phaethon was
taken from the modeling in Sec.~\ref{sec:shape}. Similarly, the volume-equivalent size of
$5.1$~km from Sec.~\ref{sec:TPM} was used as an implicit value. The last parameter
required to compute the thermal recoil acceleration (the Yarkovsky effect) is
the bulk density of Phaethon. Our nominal models use $1$ g~cm$^{-3}$ for the clarity, but we treated this value as a free parameter in the orbit determination
process \citep[similarly to what was done for asteroid Bennu in][]{Chesley2014}. Scaling to different densities is easily implemented by using the inverse-proportional dependence of the thermal acceleration
on the bulk density. In our analysis we neglected the enhancement of the Yarkovsky effect that is due to surface roughness \citep{Rozitis2012}. This effect could cause an increase in our bulk density estimate of less than  10\%, which is well within the formal uncertainty.

After determining the temperature of each surface facet, we evaluated the total thermal
acceleration at every minute throughout the orbit using a numerical surface integration.
For simplicity, we assumed the Lambert thermal emission law \citep[e.g., Eq.~(3) in][]{Bottke2006}. These thermal acceleration values were then used as part of the fit to the astrometry.

With this procedure, we estimated Phaethon's bulk density as $1.67\pm 0.47$ g~cm$^{-3}$.
Variations due to the different packing factors are negligible ($<$0.02 g~cm$^{-3}$).
Therefore, the differences stemming from a choice of the thermal model are much
smaller than the formal uncertainty of the density solution from the fit to the astrometry. We also verified that
calibrating the $\kappa(T)$ constants $\kappa_\mathrm{b}$ and $a$ to
the value of the thermal inertia $\Gamma=400$ and $800$ \tiu, that is, at one standard deviation from the nominal value, 
produces an insignificant variation in the bulk density solution
($10$\% vs. the $28$\% formal uncertainty). Similarly, our solution assumes the nominal value
of Phaethon's size, specifically $5.1$~km, which has only small fractional (formal) uncertainty
of $\simeq 4$\%. This may again be neglected with respect to the $28$\% fractional
uncertainty of the density solution. If required, we may also express the Yarkovsky
detection for Phaethon in the usual way as a $-(6.9\pm 1.9)\times 10^{-4}$
au~Myr$^{-1}$ secular drift of the orbital semimajor axis \citep[compare with data,
e.g., in][]{Vokrouhlicky2015}. 

\section{Discussion}\label{sec:discussion}

The size and albedo solution from the thermal modeling may have a systematic error
due to the possibly improper modeling of physical effects on the surface of this extreme
body. This point of view is advocated by results of the recent radar campaign
\citep[e.g.,][]{Taylor2018} and extrapolation of the polarimetric measurements at
large phase angles \citep{Ito2018,Devogele2018}, both of which yield asomewhat larger
size $\simeq (5.7-5.8)$~km than the 5.1~km nominal value used in this work. Without a detailed understanding of this difference, we
are not fully capable of correcting our basic thermal model to accommodate this
difference. If we were  to assume this larger size, we would obtain a bulk density
of $\sim$ 1.48 g~cm$^{-3}$, which is within the estimate uncertainty. We only
note that the bulk density would approach the $\simeq 1.27$ g~cm$^{-3}$ value
obtained for asteroid (101955)~Bennu \citep{Chesley2014}, which is also a small B-type body.
While this consistency is encouraging, it should not be overstated: we note, for instance,
that the suggested source region for Bennu \citep[Eulalia or new Polana asteroid
families,][]{Walsh2013,Bottke2015} is different from that of Phaethon and the spectra of these two bodies
are somewhat different in the near-infrared range \citep{Campins2010}.

Our density solution for Phaethon from the Yarkovsky model in Sec.~\ref{sec:yarko}
assumed a particular model of the temperature dependence of the surface thermal conductivity.
While supported by theoretical arguments and measurements for the lunar regolith
\citep[e.g.,][]{Keihm1984}, we do not have a direct observational confirmation of
this effect on Phaethon. On the other hand, \citet{Rozitis2018} measured a different dependence of the thermal inertia on temperature on (1036)~Ganymed, (276049) 2002 CE$_{26}$, and (1580)~Betulia. This is because the calibration of Phaethon's
surface conductivity derives from observations taken at a very restricted range of heliocentric
distances. Moreover, we note that the bulk density solution depends on the conductivity
assumption. For instance, if the conductivity were assumed constant (and not increasing
toward smaller heliocentric distances), the bulk density would decrease to $1.08\pm
0.30$ g~cm$^{-3}$. In general, any shallower dependence of the surface thermal conductivity
on heliocentric distance than that of our nominal model would imply a lower bulk density.

We assumed that the Yarkovsky effect is the dominant non-gravitational
effect in the orbital solution of Phaethon. While the effects of the Poynting-Robertson or solar wind drag are about two orders
of magnitude smaller and may be safely neglected, some concern remains
about influence of the mass loss near perihelion passages. We note that direct
observational evidence is quite limited and reveals only short episodes of very
weak activity \citep[e.g.,][]{Li2013}. Additionally, the assumed small size of the
particles triggering the observed effects results in a quite low mass-loss rate \citep[e.g.,][]{Ye2018}. 
We estimated the corresponding dynamical effects in Appendix~\ref{sec:activity} and found that the resulting change in semimajor axis of Phaethon is at most an order of magnitude smaller than the Yarkovsky effect.

Additionally, we note that the spin state solution from Sec.~\ref{sec:shape} implies
that Phaethon's pole regions are never irradiated from low zenith angles near
perihelion passages. For instance, the south rotation pole is shadowed before the
perihelion passage and becomes illuminated during and after the passage. However,
the maximum solar elevation about the local horizon at the south pole is only about
$25^\circ$ and quickly becomes even smaller within a week. The situation is
opposite for the north rotation pole. As a result, we do not expect a huge increase in
activity at the pole regions caused by the changing geometry of illumination near perihelion.


(3200) Phaethon appears to be in a cluster with asteroids (155140)~2005~UD and (225416)~1999~YC \citep{Ohtsuka2006,Ohtsuka2008,Hanus2016b}.\footnote{The 6m diameter asteroid 2012~KT$_{42}$ may be also a member of the Phaethon cluster. \citet{Polishook2012c} found it to be a B type, very similar to Phaethon.}
The cluster 3200--155140--225416 could be formed by rotational fission of a critically spinning parent body \citep[][and references therein]{Scheeres2007,Pravec2018}. To examine this hypothesis, we estimated
the total secondary-to-primary mass ratio of the cluster $q$ from the absolute magnitudes of its three members, $H_1 = 14.31, H_2 = 17.2,$ and $H_3 = 17.3$ \citep[][and MPC]{Hanus2016b} using Eqs.~(3) and (4) of \citet{Pravec2018}: $q = 0.034$.  With the primary rotation period $P_1 = 3.60$~h, this agrees excellently well with the theory of cluster formation by rotational fission. Specifically, it falls very close to the nominal $P_1$--$q$ curve in Fig.~14 of \citet{Pravec2018}, which nominally predicts $P_1 = 3.51$~h for $q = 0.034$, see Fig.~\ref{fig:pravec2018}. In other words, the current rotation of Phaethon was slowed down from the original critical spin frequency
by the formation and ejection of the two secondaries, with part of its original rotation energy and angular moment carried away by the escaping secondaries.
The apparent top-like shape of Phaethon may be a product of the spin fission process, as observed for a number of primaries of near-Earth binary asteroids. We further note that this hypothesis should not be overestimated: the uncertainty about whether the three bodies are indeed dynamically related is still great. Further physical characterization of the two smaller bodies is required to better understand the properties of the suggested cluster.

 

Our derived Phaethon bulk density is consistent with values typical for large ($>$100~km)
C-complex asteroids \citep{Carry2012b, Marchis2008, Hanus2017b}. However, similarly sized C-complex
asteroids should have higher porosity and therefore a slightly lower density than found here
for Phaethon. We may only speculate that the extreme solar irradiation is capable of
decreasing the macroporosity. Interestingly, the bulk density of Pallas is higher than
the typical values for the $D>$100~km C-complex asteroids, so that the possible dynamical link
with Phaethon \citep{deLeon2010, Todorovic2018} is consistent with Phaethon's higher bulk
density. Conversely, typical comets are found to have bulk densities far lower
\citep[often lower than $1$ g~cm$^{-3}$, e.g.,][]{Weissman2008}. Our results thus speak
against Phaethon being a comet in its nearly dormant phase.

\begin{acknowledgements}
 This work has been supported by the Czech Science Foundation (JH grant 18-04514J,
 DV and the Ond\v{r}ejov group grant 17-00774S). Additional support from the Charles University
 Research program No.~UNCE/SCI/023 is also acknowledged. MD acknowledges support from the French Centre National d’\'Etudes Spatiales (CNES). DP acknowledges the Koshland Foundation’s support. Part of the research was carried out at the Jet Propulsion Laboratory, California Institute of Technology, under a contract with the National Aeronautics and Space Administration. We thank J.~Giorgini for providing useful information about the radar astrometry of Phaethon.
\end{acknowledgements}

\bibliography{mybib}
\bibliographystyle{aa}

\begin{appendix}

\section{Optical light curves}\label{sec:lcs}

We downloaded 55 optical light curves from the DAMIT%
\footnote{\url{http://astro.troja.mff.cuni.cz/projects/asteroids3D}}
database \citep{Durech2010} that have been already used for the shape model
determination in \citet{Hanus2016b}. Moreover, we enhanced this dataset
by adding four light curves obtained in 2016 by \citet{Warner2017}, four
light curves by David Polishook (apparitions in 2005, 2007 and 2017), and
finally, four light curves obtained by Brian Warner, two by Robert Stephens
and one by Petr Pravec during the most recent apparition in December 2017.
We note that our new data sample three additional apparitions in 2005, 2016
and 2017. New observations are summarized in Table~\ref{tab:lcs}. Additional
details about the image/data processing can be found in \citet{Polishook2009}.

\begin{table*}
\caption{\label{tab:lcs}New optical photometry used to revise the shape model. The table gives the epoch, the number of individual measurements $N_p$, the asteroid distances to the Earth $\Delta$ and the Sun $r$, the phase angle $\varphi$, the photometric filter, and the observational log.}
\centering
\begin{tabular}{rlr rrr r rrr}
\hline 
\multicolumn{1}{c} {N} & \multicolumn{1}{c} {Epoch} & \multicolumn{1}{c} {$N_p$} & \multicolumn{1}{c} {$\Delta$} & \multicolumn{1}{c} {$r$} & \multicolumn{1}{c} {$\varphi$} & \multicolumn{1}{c} {Filter} & Site & Observer  & Reference \\
 &  &  & [a.u.] & [a.u.] & [deg] &  &  &  &  \\
\hline\hline
1   &  2005-11-27.0  &  62   &  1.51  &  2.40  &  12.6  &  R   &  WISE  &  David Polishook  &  This work                       \\
2   &  2007-12-02.0  &  93   &  0.21  &  1.07  &  62.2  &  R   &  WISE  &  David Polishook  &  This work                       \\
3   &  2016-11-02.2  &  49   &  0.69  &  1.49  &  34.0  &  V   &  CS3-PDS  &  Brian Warner  &  \citet{Warner2017}                            \\
4   &  2016-11-03.2  &  62   &  0.71  &  1.50  &  33.8  &  V   &  CS3-PDS  &  Brian Warner  &  \citet{Warner2017}                            \\
5   &  2016-11-04.2  &  119  &  0.72  &  1.51  &  33.5  &  V   &  CS3-PDS  &  Brian Warner  &  \citet{Warner2017}                            \\
6   &  2016-11-05.2  &  109  &  0.74  &  1.52  &  33.4  &  V   &  CS3-PDS  &  Brian Warner  &  \citet{Warner2017}                            \\
7   &  2017-11-12.1  &  100  &  0.69  &  1.49  &  33.1  &  R   &  WISE  &  David Polishook  &  This work                         \\
8   &  2017-11-23.2  &  160  &  0.46  &  1.36  &  30.4  &  R   &  D65   &  Petr Pravec, Hana Ku\v c\'akov\'a   &  This work  \\
    &                &       &        &        &        &      &        &  Kamil Hornoch, Peter Ku\v snir\'ak  &             \\
9   &  2017-11-26.3  &  89   &  0.40  &  1.32  &  29.1  &  V   &  CS3-TRJ  & Robert Stephens &  This work                            \\
10  &  2017-12-01.3  &  25   &  0.30  &  1.25  &  26.2  &  V   &  CS3-TRJ  & Robert Stephens &  This work                            \\
11  &  2017-12-01.5  &  24   &  0.30  &  1.25  &  26.1  &  V   &  CS3-PDS  &  Brian Warner  &  This work                            \\
12  &  2017-12-02.3  &  21   &  0.28  &  1.23  &  25.5  &  V   &  CS3-PDS  &  Brian Warner  &  This work                            \\
13  &  2017-12-02.4  &  12   &  0.28  &  1.23  &  25.4  &  V   &  CS3-PDS  &  Brian Warner  &  This work                            \\
14  &  2017-12-02.5  &  23   &  0.28  &  1.23  &  25.4  &  V   &  CS3-PDS  &  Brian Warner  &  This work                            \\
15  &  2017-12-09.0  &  58   &  0.16  &  1.14  &  19.7  &  R   &  WISE  &  David Polishook  &  This work                        \\
\hline
\end{tabular}
\tablefoot{
     WISE - Wise Observatory, Israel, CS3-PDS -- Center for Solar System Studies, 446 Sycamore Ave., Eaton, CO 80615, USA, D65 -- 65cm telescope at Ond\v rejov Observatory, Czech Republic, CS3-TRJ - Center for Solar System Studies, 11355 Mount Johnson Ct., Rancho Cucamonga, CA 91737, USA.
    }
\end{table*}

\section{Acceleration due to mass loss}\label{sec:activity}

Here we estimate the change in semimajor axis of Phaethon that is due to mass loss. The documented
activity of Phaethon \citep[e.g.,][]{Jewitt2010,Jewitt2013,Li2013} is
very tiny and restricted to a very narrow interval of time around
perihelion passage. \citet{Jewitt2013} mention the 2009 and 2012 events
as $\sim$2 d activity with an average mass loss rate d$M$/d$t$ $\sim$3 kg\,s$^{-1}$ and
characteristic ejection speeds between $V$ $\sim$ 10--30 m\,s$^{-1}$ (the upper value
being an order of magnitude higher than the escape speed from
Phaethon, perhaps consistent with small size of the observed particles and radiative striping from the body, rather than a traditional
jet-like activity). We note that the 2016 perihelion passage activity was
even smaller \citep{Hui2017}. With these numbers we obtain an estimate
of the effective recoil acceleration

\begin{equation}
 a_\mathrm{rec} \sim V (\mathrm{d}M/\mathrm{d}t) / M_\mathrm{3200} \sim 6.4 \times 10^{-14} \mathrm{au/d}^2.
\end{equation}
Here, $M_\mathrm{3200}$ is the estimated mass of the body, conservatively assuming
a smaller size of 5.1 km and only 1 g\,cm$^{-3}$ bulk density.

We denote $T_\mathrm{act}$ the time interval of activity around perihelion
and assume $T_\mathrm{act} \sim$ 5 d, again longer than observed so far. For the sake of
estimating the dynamical effect, we take now the most extreme possibility
that all the observed particles are ejected in a narrow jet
emanating from the north pole of Phaethon. Then the effective
orbit-averaged change in semimajor axis is

\begin{equation}
(\mathrm{d}a/\mathrm{d}t)_\mathrm{eff} \sim \frac{1}{\pi} \sqrt{((1+e)/(1-e))} T_\mathrm{act} a_\mathrm{rec} f,
\end{equation}
where the factor $\sqrt{((1+e)/(1-e))}$ stems from expressing the
perihelion velocity and the coefficient $f$ is a projection
factor of the recoil acceleration to the perihelion velocity
vector direction. With our determined pole orientation we find
$f \sim \cos(70 \,\mathrm{deg}) \sim 1/3$. By combining this, we obtain

\begin{equation}
(\mathrm{d}a/\mathrm{d}t)_\mathrm{eff} \sim 0.54 \times 10^{-4} \mathrm{au/Myr}.
\end{equation}

This is less then 10\% of our found orbital decay of
--(6.9$\pm$1.9) $\times$ $10^{-4}$ au/Myr.

We note that our assumptions were rather conservative. If the
activity comes from lower latitudes at the body, another decrease of
the effect should be expected. This is because the equatorial projection
of the recoil will be averaged by fast rotation of the asteroid.

We admit that the observed activity is mainly in very small particles. There is no direct observational evidence of a possible ejection of
larger particles during the recent perihelion passages. With this lack of
observational constraints, it is hard to say anything about this
component. It may appear that properties of the associated Geminid
stream would be a guidance. We note, however, that studies of the Geminid
activity \citep[e.g.,][]{Jakubik2015}, while indicating a rather
young age not exceeding 1000 yr, cannot be directly taken as
support for an equivalent activity within the last 30 yr of the
orbital data.

\section{Additional figures}

\begin{figure}
 \begin{center}
  \resizebox{\hsize}{!}{\includegraphics{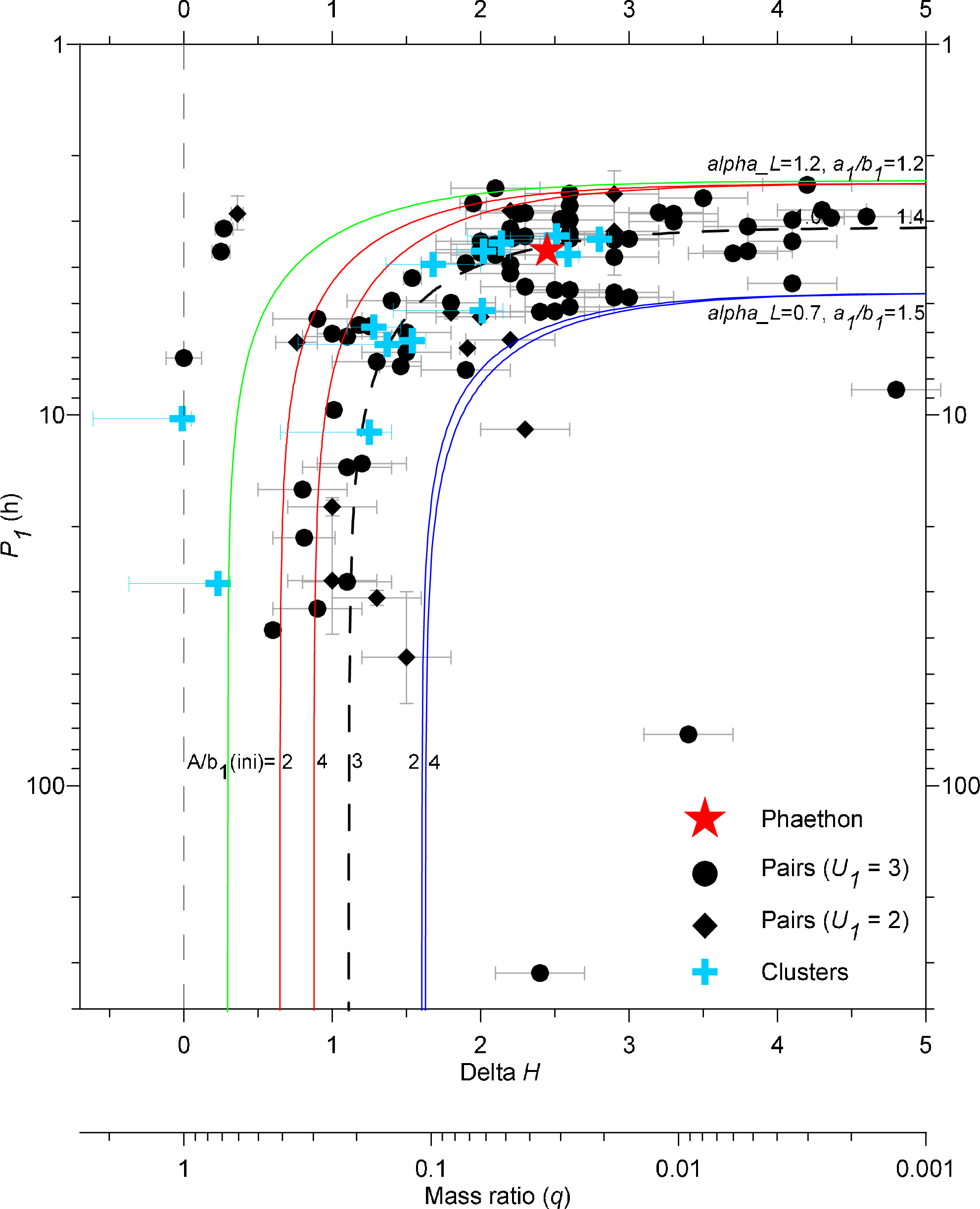}}\\
 \end{center}
 \caption{\label{fig:pravec2018} Distribution of primary rotation periods $P_1$ vs. the total secondary-to-primary mass ratios $q$ for 13 asteroid clusters, 93 asteroid pairs, and Phaethon. The figure is adopted from \citet{Pravec2018} (Fig. 14), see that paper for a full description of the figure content.}
\end{figure}

\end{appendix}

\end{document}